\documentclass[prd,reprint,superscriptaddress,floatfix]{revtex4-2}
\usepackage{soul}
\usepackage{booktabs}
\usepackage{comment}

\usepackage{color}
\usepackage{graphicx}
\usepackage{subcaption}
\usepackage{amsmath}
\usepackage{array}
\usepackage{float}
\usepackage{lineno}
\usepackage[normalem]{ulem}
\usepackage{siunitx}
\usepackage[bottom]{footmisc}
\usepackage{ragged2e}
\usepackage[dvipsnames]{xcolor}
\definecolor{apsblue}{HTML}{2E3092}
\usepackage[colorlinks=true,
            linkcolor=apsblue,
            citecolor=apsblue,
            urlcolor=apsblue]{hyperref}

\DeclareCaptionJustification{justified}{\justifying}

\begin{document}

\title{Constraints on Complex Electroweak WIMPs from Solar Capture}

\def\um{William I. Fine Theoretical Physics Institute, University of Minnesota, MN 55455, United States}
\def\cern{Theoretical Physics Department, CERN, 1 Esplanade des Particules, CH-1211 Geneva 23, Switzerland}
\def\udelbartol{Department of Physics and Astronomy, University of Delaware and the Bartol Research Institute, Newark, DE 19716, USA}

\author{Juan P. Luengas}\email{jluengas@udel.edu}\affiliation{\udelbartol}
\author{Maxim Pospelov}\affiliation{\um}\affiliation{\cern}
\author{Harikrishnan Ramani}\affiliation{\udelbartol}

\date{\today}
\begin{abstract}
One of the most convincing realizations of particle dark matter is provided by models based on electroweak multiplets. The lightest neutral state of such a multiplet can be a stable Majorana fermion or a real scalar, and its phenomenology is governed by two parameters: the dark matter mass and the mass splitting $\{m_\chi,\delta\}$. The sub-MeV range for $\delta$ is particularly important as it allows for a large inelastic scattering cross section. In this paper, we analyze solar capture, energy loss, and annihilation of scalar and fermionic electroweak multiplet dark matter, generalizing earlier results for the Higgsino to other representations. We find that, while the case of a doublet is somewhat special, higher representations behave in a self-similar way. For the higher multiplets, IceCube results constrain the splitting to $\delta \gtrsim350$ keV for all dark matter masses up to $100$ TeV. Assuming thermal freeze-out and a Standard Halo Model velocity distribution, this rules out most models of electroweak dark matter as a possible origin of the single high-recoil event reported by the LZ collaboration. We also show that a high-velocity tail in the dark matter distribution, such as that induced by the Large Magellanic Cloud, raises the splitting preferred by the LZ event above the solar bounds for many of the multiplets.
\end{abstract}

\maketitle

\section{Introduction}

Electroweak multiplets are among the simplest and most predictive candidates
for particle dark matter. The dark matter particle is the lightest neutral
component of a single multiplet of $SU(2)_L\times U(1)_Y$, interacting with
the Standard Model through the electroweak gauge interactions.
Once the spin and electroweak quantum numbers of the multiplet are specified,
requiring thermal freeze-out to reproduce the observed dark matter abundance
selects a characteristic mass, from about a TeV for the doublet to hundreds of TeV
for the largest representations consistent with perturbative
unitarity~\cite{Cirelli:2005uq,Bottaro:2022one}. The familiar supersymmetry-motivated Wino and Higgsino
are the simplest examples of a much broader class of weakly interacting massive particles (WIMPs).

The experimental phenomenology depends qualitatively on the hypercharge of
the multiplet. For $Y=0$, the neutral state has no tree-level coupling to
the $Z$ boson and its elastic scattering from nuclei first arises at loop
level. For $Y\neq0$, by contrast, a complex electroweak multiplet contains
a neutral Dirac state with a tree-level vector coupling to the $Z$. The vector coupling gives rise to spin-independent scattering on nuclei with coherent enhancement for heavy atomic species. If this
interaction were unsuppressed, the resulting nuclear scattering cross section
would exceed current direct detection limits by several orders of magnitude.
A viable dark matter candidate therefore requires the Dirac state to be
split into two Majorana states, converting the $Z$ interaction into an
inelastic transition. Such a splitting arises from electroweak symmetry breaking. In supersymmetric models it results from the mixing of Higgsino states with supersymmetric partners of $SU(2)$ and $U(1)$ gauge bosons, but similar mechanisms can be invoked for higher multiplets as well. In the language of effective field theory, the splitting is generated through 
higher dimensional operators~\cite{Bottaro:2022one} when states heavier than the electroweak multiplet are integrated out. 

The neutral splitting $\delta$ is the central parameter
governing the observability of these WIMPs, and in the absence of strong guidance for the UV physics, it makes perfect sense to keep $\delta$ as a free parameter constrained (or determined) by experiment. The splitting leaves the relic abundance
unmodified but can suppress and even switch off the WIMP-nucleus scattering rate completely \cite{Han:1997wn,Hall:1997ah,Tucker-Smith:2001myb}. Increasing $\delta$ eventually makes the
inelastic transition kinematically inaccessible. For scattering from a
nucleus of mass $m_A$, the transition requires
\begin{equation}
    \delta \lesssim \frac{1}{2}\mu_A v^2,
    \label{eq:kin}
\end{equation}
where $\mu_A$ is the dark matter--nucleus reduced mass and $v$ is their
relative velocity. Terrestrial searches therefore encounter a kinematic
limitation set by the finite velocity of dark matter in the Galactic halo:
for xenon it corresponds to $\delta\simeq 370$--$390$~keV, nearly independent
of the dark matter mass above a TeV. Once the splitting exceeds this
threshold, increasing the detector exposure cannot recover sensitivity.
Just below it, only the fastest particles in the halo contribute, and the
signal is a small number of recoils at energies of a few hundred keV, outside
the standard WIMP search window. The single $248$~keV nuclear recoil recently
reported by LZ in an extended-energy search~\cite{LZ:2026axp} falls in
precisely this regime, and electroweak WIMPs with $\delta$ just below the
xenon threshold were among its first
interpretations~\cite{Fan:2026kxx,Freese:2026sga,Wu:2026nhi,Langhoff:2026ujr,Smirnov:2026aqk,Yin:2026jnn,Du:2026guj,Bisal:2026khf,Khan:2026zuj,Chatterjee:2026scv,Cheung:2026byg,Kotlarski:2026pep,Maier:2026qpr,Nomura:2026qyq,DiMauro:2026ldr,Murayama:2026apt}.
The event has also been interpreted in a wide range of other scenarios, from
inelastic dark matter with dark-sector mediators and exothermic or composite
dark matter to boosted particles and non-dark-matter
explanations~\cite{Yamashita:2026ump,McCabe:2026crm,Jeesun:2026vzo,Unwin:2026rdp,deLima:2026shq,Gu:2026vto,Baer:2026fpy,Lee:2026wof,Liang:2026coz,Das:2026uyy,Alhazmi:2026efz,Okada:2026eol,Du:2026lpa,Kannike:2026qyl,Yuan:2026djt,Zhu:2026dag,Aghaie:2026vsu,Asadi:2026iot,Khan:2026nwp,He:2026hqz,Fan:2026hzw,Chattaraj:2026fxn,Kumar:2026lgi,Heikinheimo:2026kwp,Lee:2026zbr,Mahapatra:2026glu,Barman:2026omh,Das:2026buc,Ahmed:2026kan,He:2026idw,Lian:2026hpm,Baer:2026yrt,Uttayarat:2026isp,Arcadi:2026kev,An:2026pkc,Xing:2026civ,Okada:2026fef,Ahmed:2026com,De:2026win,Sheng:2026tqt,Chauhan:2026udz,Sannino:2026hkc,Khan:2026osp,Jung:2026otm,Gemmell:2026yaw,Tang:2026lgu,Su:2026rwz,Palmisano:2026kuj,Ge:2026xax,Lee:2026xxh,Le-Yaouanc:2026djt,Yang:2026wpb,Brdar:2026ukx,Kalita:2026hvb,Qi:2026vyp}. More data collected by the dark matter experiments will clarify whether this finding is reinforced by further signal events or traced to a new source of background. 

The Sun provides a natural way to overcome terrestrial limitations and extend sensitivity to higher $\delta$. Dark matter
falling into the solar gravitational potential is accelerated to velocities
exceeding $1300~{\rm km\,s^{-1}}$ in the solar interior. Moreover, although
heavy elements constitute only a small fraction of the solar composition,
their larger reduced mass in Eq.~\eqref{eq:kin} makes them particularly
effective targets for endothermic scattering. The combination of large
infall velocities and heavy nuclear targets allows the Sun to probe neutral
splittings far beyond those accessible to terrestrial direct detection. An
electroweak WIMP that loses sufficient energy in such a collision becomes
gravitationally bound to the Sun. Repeated scattering can subsequently
drive the captured population toward the solar core, where dark matter
annihilation produces energetic Standard Model particles and, ultimately,
high-energy neutrinos that escape the Sun and can be searched for by
neutrino telescopes~\cite{IceCube:2025fcu} via charged-current conversion of $\nu_\mu$ and $\bar\nu_\mu$. This strategy was first applied
to inelastic dark matter models~\cite{Nussinov:2009ft,Menon:2009qj} motivated by
the DAMA anomaly \cite{DAMA:2008jlt}. In Ref.~\cite{Pospelov:2026ewn} two of us
applied it to the thermal Higgsino and found that the IceCube limit excludes
$\delta\lesssim 566$~keV, well beyond the reach of xenon and, for standard
halo parameters, above the splitting that would explain the LZ event; this
conclusion has since been confirmed by several
groups~\cite{Nguyen:2026lui,Bose:2026ndd,Bose:2026szs,DiMauro:2026dqp,Ghosh:2026txe},
and proposals to evade it include singlet--doublet
mixing~\cite{Lee:2026jxl,Nagata:2026pbj}, an environment-dependent
splitting~\cite{Girmohanta:2026qzo} and non-thermal or heavier
Higgsinos~\cite{Langhoff:2026ujr,Unwin:2026qep}.

In this work, we derive solar constraints on inelastic electroweak WIMPs
across the complex representations of Ref.~\cite{Bottaro:2022one}, for both
fermionic and scalar multiplets.
We show that solar capture and subsequent annihilation probe a
region of electroweak WIMP parameter space that is inaccessible to
conventional direct detection, and that the higher multiplets can
explain the LZ event while evading the solar bounds.

\section{Electroweak multiplet Dark Matter}

\label{sec:ewdm}

We consider dark matter that is the lightest neutral component of a single
$SU(2)_L$ multiplet $\chi$ of dimension $n$ and hypercharge $Y\neq0$, with
electric charges $Q=T_3+Y$. Requiring a neutral component fixes $Y$ to be
half-integer for even $n$ and integer for odd $n$, and requiring perturbative unitarity and a viable neutral splitting restricts the
list to $n\leq 12$ for $Y=1/2$ and $n\leq 5$ for $Y=1$~\cite{Bottaro:2022one}.
We follow the classification and notation of Ref.~\cite{Bottaro:2022one} and
denote a multiplet by $n_Y$; the Higgsino is the $2_{1/2}$. Inelastic electroweak multiplets beyond the doublet, and singlet--doublet
mixtures, have been considered in the context of the LZ event in
Refs.~\cite{Smirnov:2026aqk,DiMauro:2026ymt,Visinelli:2026kgt,Wang:2026ytg,Bandyopadhyay:2026gjw,Borah:2026zwf,Borah:2026ris,Elahi:2026vlm,Okada:2026upm,Paul:2026okh,Capolupo:2026uyi,Ahmed:2026qjg,Chattopadhyay:2026ryw,Langhoff:2026qgo}. Both Dirac fermions
and complex scalars are considered. For definiteness we write the fermionic case
below and note the scalar differences where they arise.

\subsection{Neutral splitting}
The renormalizable Lagrangian couples the neutral component to the $Z$ through
its hypercharge,
\begin{equation}
\mathcal{L}_Z=\frac{g\,Y}{c_W}\,\bar\chi_N\gamma^\mu\chi_N Z_\mu ,
\label{eq:LZcoupling}
\end{equation}
which for a Dirac state gives an elastic spin-independent cross section on
nuclei several orders of magnitude above present bounds. The minimal remedy is
the operator~\cite{Bottaro:2022one}
\begin{equation}
\mathcal{O}_0=\frac{1}{2(4Y)!}\,\bigl(\bar\chi\,(T^a)^{2Y}\chi^c\bigr)
\Bigl[(H^{c})^\dagger\frac{\sigma^a}{2}H\Bigr]^{2Y}+{\rm h.c.},
\label{eq:O0}
\end{equation}
which after electroweak symmetry breaking generates a
Majorana mass for the neutral component. The neutral Dirac state splits into two Majorana states,
$\chi_1$ and $\chi_2$, with masses $m_\chi\mp\delta/2$, and the vector current in
Eq.~\eqref{eq:LZcoupling} becomes purely off-diagonal,
\begin{equation}
\mathcal{L}_Z=\frac{g\,Y}{c_W}\,\bar\chi_2\gamma^\mu\chi_1 Z_\mu+{\rm h.c.}
\label{eq:Zinel}
\end{equation}
Elastic scattering through the $Z$ is thereby removed at tree level and only the
endothermic transition $\chi_1 A\to\chi_2 A$ survives, with the amplitude fixed
entirely by $Y$. 

The differential cross section for the endothermic transition on a nucleus with
$N_A$ neutrons and $Z_A$ protons is~\cite{Pospelov:2026ewn}
\begin{equation}
\frac{d\sigma_A}{dE_R}=\frac{G_F^2\,m_A\,Y^2}{\pi\,v^2}
\Bigl[N_A-\bigl(1-4\sin^2\theta_W\bigr)Z_A\Bigr]^2
F_A^2,
\label{eq:dsigma}
\end{equation}
where $v$ is the dark matter velocity and $F_A$ the
nuclear form factor. This is the interaction responsible for capture in the Sun. For
a complex scalar the same construction applies with $\mathcal{O}_0$ built from
$\chi^2 H^{4Y}$, and Eq.~\eqref{eq:Zinel} is replaced by the analogous
off-diagonal derivative coupling; the resulting nuclear cross section is
identical to the fermionic one in the non-relativistic limit. Scalar electroweak WIMPs have one important distinction from fermions. The dimension-four operator $(\chi^*\chi)(H^\dagger H)$ is always allowed, and from the EFT perspective is not suppressed by an ultraviolet scale. If this term enters with a sizable coefficient, it can regulate the primordial WIMP abundance, even for a completely neutral WIMP \cite{Silveira:1985rk,McDonald:1993ex,Burgess:2000yq}. Higgs exchange mediates elastic scattering at tree level, which may be problematic from the point of view of current direct detection bounds. For simplicity, we will assume that such an interaction is not present at tree level and is induced only via loops.

In our paper, we will also limit the range of $Y\leq 1$. It was pointed out in Ref.~\cite{Bottaro:2022one} that larger $Y$ raises the dimension of the operator~(\ref{eq:O0}), and that the resulting suppression by the natural UV scale, $\Lambda_{\rm UV}\gtrsim m_\chi$, then makes $\delta$ too small. While this is generically true, there can be some spectra of electroweakly charged particles that would defy this logic. In particular, if the lowest mass WIMP state {\it coexists} with electroweak states with smaller $Y$ and slightly larger mass, their effective mixing will be suppressed by the scale of $\Delta m$, which can be much smaller than $m_\chi$, thereby allowing sizable $\delta$. For simplicity, we disregard this caveat, and take $Y\leq 1$.

\subsection{Annihilation and thermal masses}
\label{sec:annihilation}

\begin{figure*}[!t]
    \centering

    \includegraphics[width=0.95\textwidth]{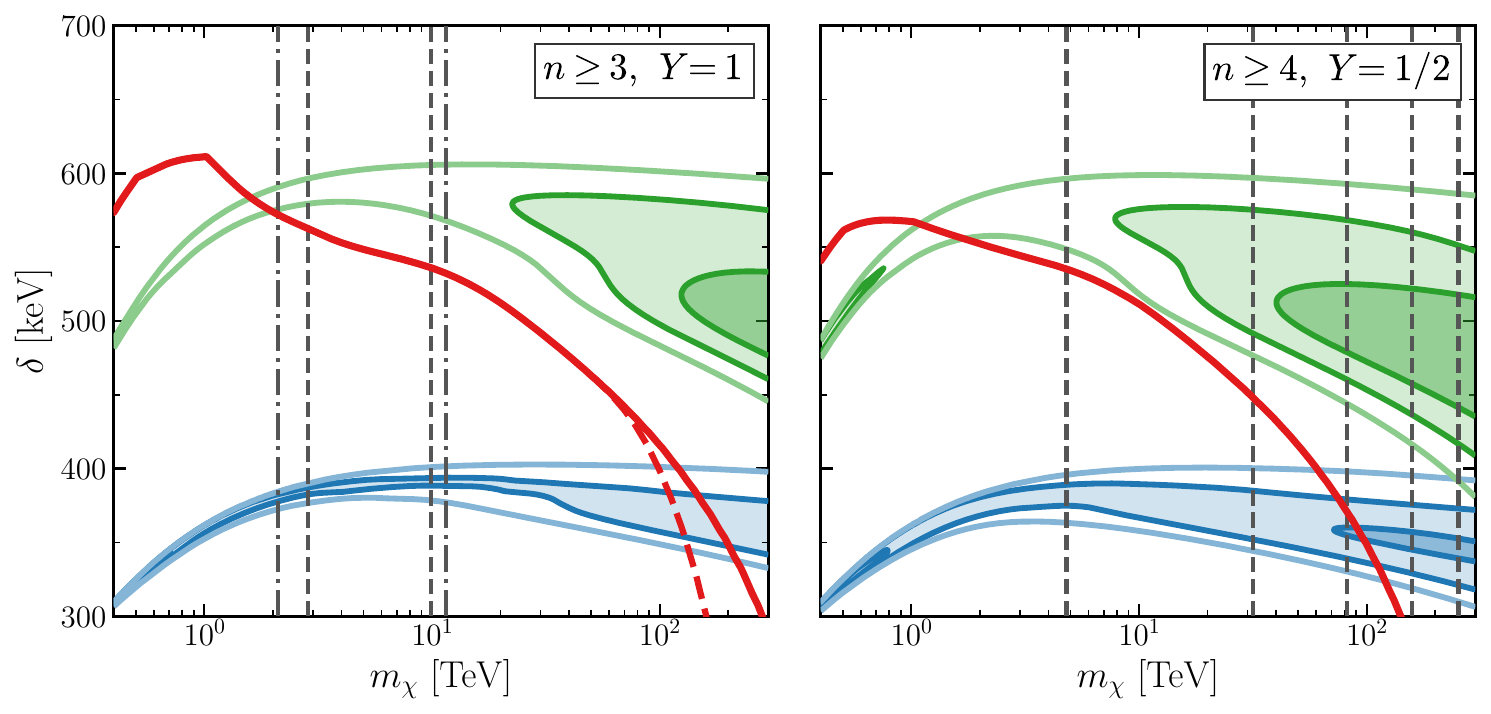}

    \caption{Mass-splitting maps in the $(m_\chi,\delta)$ plane for the electroweak multiplets with hypercharge $Y=1$ (left panel) and $Y=1/2$ (right panel). Blue and green contours correspond to the SHM and LMC $1\sigma$, $2\sigma$, and $3\sigma$ regions, from dark to light, respectively. Solid red curves show the solar constraints derived in Sec.~\ref{sec:results}, while the dashed red curve corresponds to the \(3_1\) fermion multiplet, which is not in capture–annihilation equilibrium at large masses. Its scalar counterpart is not shown for clarity, differing by a factor $\sqrt2$ through the annihilation rate and are otherwise identical. The vertical dashed and dash-dotted lines indicate the relic masses of the fermionic and scalar multiplets. For $Y=1/2$, only fermionic relic masses are shown, as the scalar ones differ by just a few TeV.}

    \label{fig:npletsmap}
\end{figure*}

\begin{figure*}[!t]
    \centering

    \includegraphics[width=0.95\textwidth]{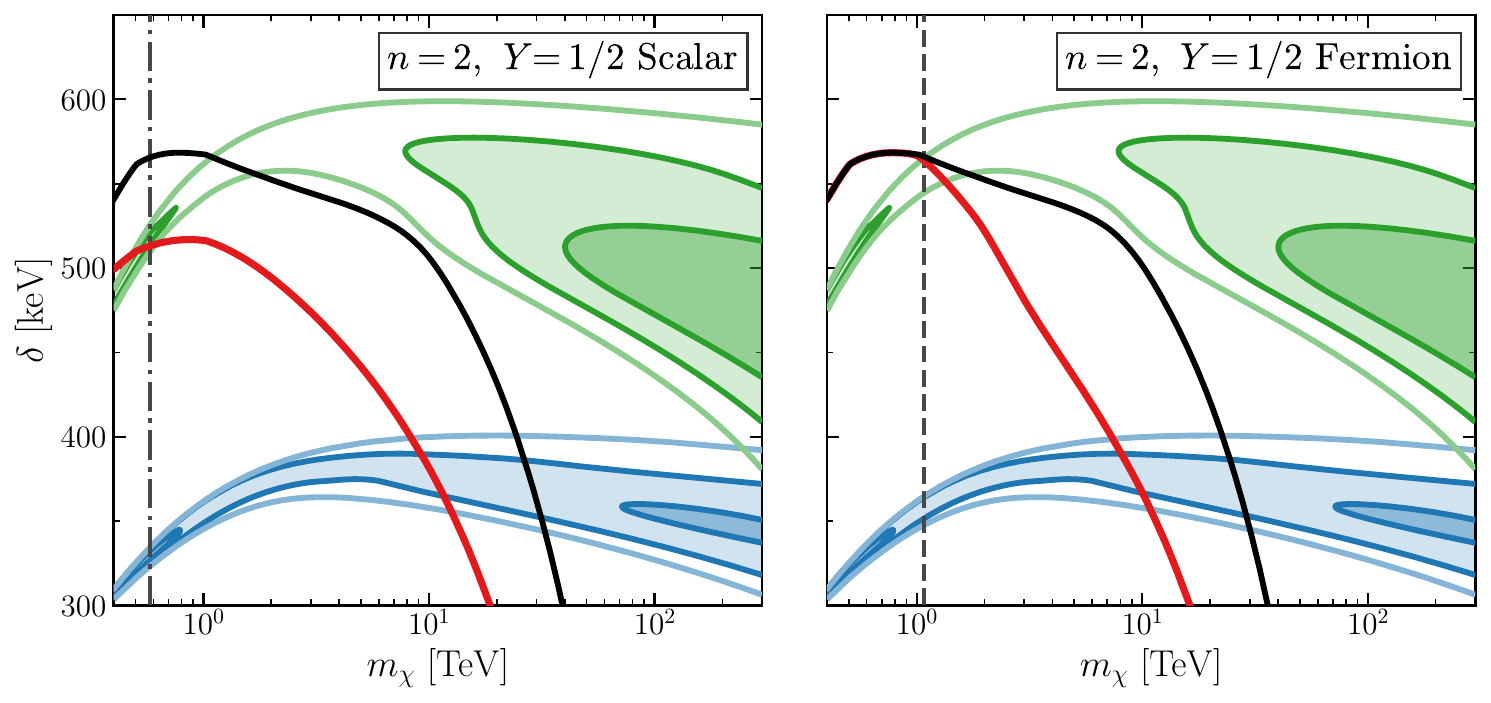}

    \caption{Mass-splitting maps in the $(m_\chi,\delta)$ plane for the scalar doublet (left panel) and the fermionic doublet (right panel). The blue contours correspond to the SHM halo model, while the green contours correspond to the LMC halo $1\sigma$, $2\sigma$, and $3\sigma$ regions, from dark to light. The red curves show the solar constraints for each case, derived in Sec.~\ref{sec:results} for $\sigma_{\rm SI}^{\rm loop}=0$ in red and taking $\sigma_{\rm SI}^{\rm loop}=4\times10^{-50}\,\mathrm{cm^2}$, its central value from Ref.~\cite{Hill_2014} in black. The region between the red and the black curves hence acts as the uncertainty band in the solar limits arising from the poorly known $\sigma_{\rm SI}^{\rm loop}$. The vertical dashed/dash-dotted lines indicate the corresponding relic masses.}

    \label{fig:dirac_nplets_shm_lmc}
\end{figure*}

At low velocity the neutral state annihilates into $W^+W^-$ and $ZZ$ through
$t$-channel exchange of the charged and neutral partners, with tree-level rate
\begin{align}
\langle\sigma v\rangle^{\rm tree}_F
&=\frac{\pi\alpha_2^2}{m_\chi^2}\left[\frac12\left(\frac{n^2-1}{4}-Y^2\right)^2
+\frac{Y^4}{c_W^4}\right],\nonumber \\
\langle\sigma v\rangle^{\rm tree}_S&=2\,\langle\sigma v\rangle^{\rm tree}_F ,
\label{eq:svtree}
\end{align}
where the two terms correspond to the $W^+W^-$ and $ZZ$ final states and the
factor of two for scalars is due to the contact term. The long-range electroweak
potential enhances this rate (Sommerfeld enhancement) and, for the larger
multiplets, adds bound-state
formation~\cite{Hisano:2004pv,Beneke:2014hja,Bottaro:2022one}. This is especially relevant for annihilation in the Galaxy or in the Sun, where generically $v/c < 10^{-3}$. For the primordial annihilation at the freeze-out, $v/c\sim 0.3$ and the Sommerfeld factor is less important. 

The annihilation rate grows with $n$; its resonant structure depends on the mass and
on the neutral and charged splittings, which set the thresholds of the
coupled two-body channels, and for $n\ge4$ the charged splitting is a free
UV parameter. 
In order to be free from specific model choices, we estimate the
enhancement parametrically, from the attractive potential in the
non-relativistic, off-resonance limit following Ref.~\cite{Zavala:2009mi},
\begin{equation}
S\simeq 6\,|\alpha_{\rm eff}^{I}|\,\frac{m_\chi}{m_W},
\qquad
\alpha_{\rm eff}^{I}=\frac{\alpha_2}{8}\bigl(I^2+1-2n^2\bigr)+Y^2\alpha_Y ,
\label{eq:sest}
\end{equation}
where $I$ is the dimension of the isospin representation of the annihilating
pair~\cite{Bottaro:2022one} and $\alpha_{\rm eff}<0$ is attractive. For smaller masses, this expression is replaced by $S\sim1$. Conversely, at very large masses such that $m_\chi v \geq  m_W$, the enhancement is replaced with $S\sim2\pi|\alpha_{\rm eff}|/v $, where $v$ is the characteristic velocity. The most
attractive channel is the singlet, $I=1$, which dominates the low-velocity
annihilation, and we use it throughout. For the doublet,
$|\alpha^{I=1}_{\rm eff}|=0.023$ and $S\simeq1.7\,(m_\chi/{\rm TeV})$,
compared with $1.77$ at $1.08$~TeV obtained from the full numerical coupled-channel calculation including the
 radiative charged splitting.
 More generally, we have checked Eq.~\eqref{eq:sest} against
the corresponding numerical calculation for the $2_{1/2}$, $3_1$ and $5_1$ over masses from $1$--$300$~TeV. Away from resonances, it
tracks the exact result to within $30\%$, while near them the exact result is larger. The resulting Sommerfeld-enhanced rate,
$\langle\sigma v\rangle^{\rm SE}=S\,\langle\sigma v\rangle^{\rm tree}\propto1/m_\chi$,
is listed in Table~\ref{tab:nplets}. Stepping ahead, the Sommerfeld-enhanced rates are relevant for the solar capture and annihilation problem, when the dynamical equilibrium has not been reached.

Requiring freeze-out to reproduce $\Omega h^2=0.12$ fixes the mass of each
representation. Since $\langle\sigma v\rangle$ grows as $n^4/m_\chi^2$, the
thermal mass grows roughly as $n^2$, from $1.08$~TeV for the doublet to
$253$~TeV for the $12_{1/2}$; for $n\gtrsim4$ the Sommerfeld and bound-state
effects raise it by a further factor of two to three over the tree-level
estimate. We adopt the masses computed with these effects in
Ref.~\cite{Bottaro:2022one}, listed in Table~\ref{tab:nplets} for Dirac
fermions and complex scalars together with the loop cross section of
Eq.~\eqref{eq:sigmaSI_loop} and the coefficient of Eq.~\eqref{eq:svtree}. The
quoted uncertainties are at the $5$--$15\%$ level and do not affect our
conclusions. The scalar masses are close to the fermionic ones for $n\geq3$;
the scalar doublet is the exception, at $0.58$~TeV. The thermal mass is the
natural benchmark for each multiplet and is marked in every figure, but we
present the constraints over the full range $m_\chi\lesssim300$~TeV, since a
non-standard cosmology can place the
mass elsewhere. In total, there are 16 thermal relic targets across $Y=\frac{1}{2}$ and $Y=1$ and across scalars and fermions. 

\begin{table*}[t]
\centering
\begin{tabular}{c cc c cc}
\hline\hline
 & \multicolumn{2}{c}{$M_{\rm th}$ [TeV]} & & \multicolumn{2}{c}{$\langle\sigma v\rangle^{\rm SE}\,(m_\chi/{\rm TeV})$ [cm$^3$\,s$^{-1}$]} \\
\cmidrule(lr){2-3}\cmidrule(lr){5-6}
multiplet & fermion & scalar & $\sigma_{\rm SI}^{\rm loop}$ [cm$^2$] & fermion & scalar \\
\hline
$2_{1/2}$  & $1.08\pm0.02$ & $0.58\pm0.01$ & $\lesssim10^{-49}$ & $1.6\times10^{-26}$ & $3.3\times10^{-26}$ \\
$3_{1}$    & $2.85\pm0.14$ & $2.1\pm0.1$   & $2.9\times10^{-49}$ & $3.9\times10^{-25}$ & $7.9\times10^{-25}$ \\
$4_{1/2}$  & $4.8\pm0.3$   & $4.98\pm0.25$ & $3.3\times10^{-47}$ & $2.4\times10^{-24}$ & $4.8\times10^{-24}$ \\
$5_{1}$    & $9.9\pm0.7$   & $11.5\pm0.8$  & $4.5\times10^{-47}$ & $8.5\times10^{-24}$ & $1.7\times10^{-23}$ \\
$6_{1/2}$  & $31.8\pm5.2$  & $32.7\pm5.3$  & $2.2\times10^{-46}$ & $3.3\times10^{-23}$ & $6.6\times10^{-23}$ \\
$8_{1/2}$  & $82\pm8$      & $84\pm8$      & $7.6\times10^{-46}$ & $2.0\times10^{-22}$ & $4.0\times10^{-22}$ \\
$10_{1/2}$ & $158\pm12$    & $162\pm13$    & $1.9\times10^{-45}$ & $7.8\times10^{-22}$ & $1.6\times10^{-21}$ \\
$12_{1/2}$ & $253\pm20$    & $263\pm22$    & $4.1\times10^{-45}$ & $2.4\times10^{-21}$ & $4.8\times10^{-21}$ \\
\hline\hline
\end{tabular}
\caption{Thermal relic masses from Ref.~\cite{Bottaro:2022one}, including
Sommerfeld enhancement and bound-state formation; the lower limit on the loop-induced elastic cross
section from Eq.~\eqref{eq:sigmaSI_loop}, identical for fermions and scalars; and the
Sommerfeld-enhanced annihilation rate
$\langle\sigma v\rangle^{\rm SE}=S\,\langle\sigma v\rangle^{\rm tree}$ with
$S=6|\alpha^{I=1}_{\rm eff}|m_\chi/m_W$ from Eqs.~\eqref{eq:svtree}
and~\eqref{eq:sest}, which scales as $1/m_\chi$; the scalar rate is twice the
fermionic one. For the doublet, the full SI calculation is consistent with zero. The fermion doublet thermalizes through the spin-dependent channel, while the scalar doublet has no SD coupling; see Sec.~\ref{sec:energyloss}.}
\label{tab:nplets}
\end{table*}
\subsection{Elastic scattering at one loop}
With the tree-level $Z$ exchange rendered inelastic, elastic scattering of
$\chi_1$ on nuclei proceeds through $W$ and $Z$ loops and Higgs exchange. In the heavy-WIMP limit the loop-induced
SI nucleon cross section is
\begin{equation}
\sigma_{\rm SI}^{\rm loop}
\simeq
\frac{4}{\pi}m_N^4
\left|k_N^{\rm EW}\right|^2 ,
\label{eq:sigmaSI_loop}
\end{equation}
where $k_N^{\rm EW}$ corresponds to the full electroweak
one-loop matching of Refs.~\cite{Bottaro:2022one, Hisano:2011cs}. For the doublet the resulting SI cross section sits near the neutrino floor. Moreover, careful higher-order calculations~\cite{Hill:2014yxa} show that the SI loop cross section is consistent with zero, hence we show solar limits for both zero as well as the central value. However, for all other multiplets Eq.~\eqref{eq:sigmaSI_loop} is a very good approximation for the loop cross section. 

The scalar cross
section is the same to the accuracy of Eq.~\eqref{eq:sigmaSI_loop}, under the additional assumption that the quartic WIMP--Higgs operator is absent at tree level. The
elastic channel is irrelevant for capture, which is dominated instead by the tree-level
inelastic transition. But it is what allows a captured
particle to keep losing energy once its velocity has dropped below the
inelastic threshold.

For the case that $\sigma_{\rm SI}^{\rm loop}$ is consistent with zero for the fermionic doublet, similarly to
Ref.~\cite{Pospelov:2026ewn}, we instead use the loop-induced spin-dependent
cross section. The axial coupling does not suffer the cancellation of the
scalar one but, unlike it, vanishes in the heavy-WIMP limit,
$d_q\propto\alpha_2^2/(m_W m_\chi)$~\cite{Hisano:2011cs}, so that the
spin-dependent cross section falls as $m_\chi^{-2}$:
\begin{equation}
\sigma_{\rm SD}^{p}\simeq 4\times10^{-46}~{\rm cm}^2
\left(\frac{1~{\rm TeV}}{m_\chi}\right)^{2}
\left(\frac{\Delta\Sigma_p}{0.32}\right)^{2},
\label{eq:sigmaSD_loop}
\end{equation}
\enlargethispage{\baselineskip}
where $\Delta\Sigma_p=\Delta u+\Delta d+\Delta s$ is the singlet axial charge
of the proton \cite{Thomas:2008bd}, which carries most of the uncertainty: older determinations
give $\Delta\Sigma_p\simeq0.13$ \cite{SpinMuon:1995svc} and a cross section an order of magnitude
 smaller, which is the value used in Ref.~\cite{Pospelov:2026ewn}. This channel acts only on the solar hydrogen, without the coherent $A^2$
enhancement of the spin-independent one, and it is what brings the captured
fermionic doublet toward the core. The scalar doublet has no spin and hence
no spin-dependent coupling at all. For $n>2$ the channel is irrelevant: the
coherent $\sigma_{\rm SI}^{\rm loop}$-induced energy loss rate on iron exceeds the spin-dependent rate
on hydrogen by many orders of magnitude.

\subsection{LZ 248 keV event}
\label{sec:lz}

The LZ collaboration has reported a single nuclear recoil at
$E_R=248\pm23_{\rm stat}\pm23_{\rm sys}$~keV in a search extending the recoil
window to $270$~keV, with an expected background of $0.01$ events in the
surrounding energy slice~\cite{LZ:2026axp}. The same data contain no events
in the high-energy sideband that the collaboration uses for background
validation, $800<{\rm S1c}<1700$~phd, corresponding to
$350\lesssim E_R\lesssim675$~keV. For an inelastic electroweak WIMP both
numbers carry information, and we fit $(m_\chi,\delta)$ to them with the signal-only likelihood of Ref.~\cite{Langhoff:2026ujr}, $-\ln\mathcal{L}=N_{\rm ROI}-\ln p(248~{\rm keV})$, where $N_{\rm ROI}$ is the expected signal count over the full recoil window up to $675~{\rm keV}$, obtained by extrapolating the LZ efficiency to high recoil energies and folding the spectrum with the LZ energy resolution, while $p$ is the corresponding event density at the observed recoil energy. The
cross section is fixed to its electroweak value, Eq.~\eqref{eq:dsigma}, so the
only freedom is in the mass and the splitting. We use two velocity
distributions: the Standard Halo Model, truncated at the Galactic escape velocity, and the LMC-motivated model of
Ref.~\cite{Smith-Orlik:2023kyl}. This reference suggests, through numerical simulations, that inclusion of LMC gravity leads to an enhanced high-velocity tail in the dark matter distribution relative to the Standard Halo Model. We will refer to this possibility as the ``LMC model'' for concision.

The resulting $1$, $2$ and $3\sigma$ regions are shown in
Figs.~\ref{fig:npletsmap} and~\ref{fig:dirac_nplets_shm_lmc}, together with
the thermal masses of the multiplets of Table~\ref{tab:nplets}. The preferred
regions agree well with those of Ref.~\cite{Langhoff:2026ujr} for both halos.
The same
figures also show, in red, the solar constraints that are the subject of the
rest of this paper; they are derived in Sec.~\ref{sec:results}, and for now the
reader may ignore them. The empty sideband, whose importance was noted
in Refs.~\cite{Rodd:2026tyn,Langhoff:2026ujr,Dent:2026bji,Delepine:2026ith}, penalizes masses near a TeV: at fixed
$\delta$ the recoil spectrum of a TeV-scale WIMP extends well past $350$~keV
and predicts several sideband events for every event in the region of
interest. The fit therefore prefers either masses below $500$~GeV, where
$m_\chi$ and $\delta$ can be adjusted independently to place the recoil at
$248$~keV with an empty sideband, or masses above $100$~TeV, where
$\mu_A\to m_A$ and the region of interest and the sideband are populated at
comparable levels~\cite{Langhoff:2026ujr}. It should be borne in mind that these regions are drawn from a single event. In the LMC model, the two islands are separated only at the $1$--$2\sigma$ level, while the $3\sigma$ contour remains continuous across the full mass range. For the SHM, the $2$--$3\sigma$ contours are continuous across all masses.

Also shown is the full catalogue of electroweak WIMPs. As mentioned earlier, WIMPs with the same hypercharge, irrespective of spin or representation, have the same direct detection cross section. This allows us to group them together in the same plot, one each for $Y=1$ and $Y=\frac{1}{2}$. The doublets are shown separately in Fig.~\ref{fig:dirac_nplets_shm_lmc} owing to subtleties in their thermalization in the Sun.

Among the thermal candidates, the scalar doublet lies within the $3\sigma$ region for both velocity distributions, whereas the fermionic doublet lies within the $2\sigma$ region for the SHM and the $3\sigma$ region for the LMC, as shown in Fig.~\ref{fig:dirac_nplets_shm_lmc}. 
The $Y=\frac{1}{2}$ multiplets (both scalars and fermions) for $n\ge 6$ lie within the high-mass $2\sigma$ island for both velocity distributions. All the other thermal multiplets are still within the $ 3\sigma$ band. Which of these thermal relic candidates survives the
solar constraint is the subject of Sec.~\ref{sec:results}.

\section{Capture and evolution in the Sun}
\label{sec:capture}
\begin{figure}[t]
    \centering

    \includegraphics[width=\columnwidth]{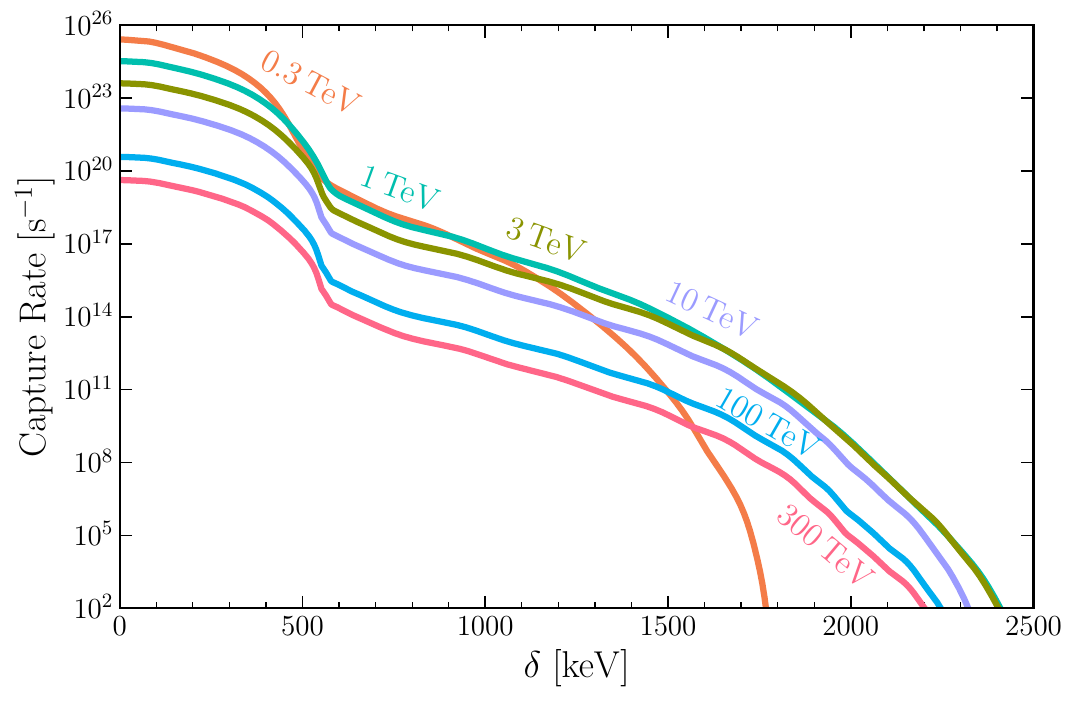}

    \caption{Total solar capture rate for $Y=\tfrac12$ electroweak multiplets as a function of the mass splitting, for several dark matter masses. The rate is the same for every $n$ and for scalars and fermions; for $Y=1$ it is four times larger.}

    \label{fig:capture_relic_masses}
\end{figure}
The inelastic scattering of
Eq.~\eqref{eq:dsigma} controls capture and is fixed by $Y$ alone, so that the
capture rate of every complex multiplet is a known multiple of the Higgsino one
at the same mass and splitting. The kinematics of this capture rate were discussed in Ref.~\cite{Pospelov:2026ewn} and will not be repeated here. We plot the capture rate $\mathcal{C}$ as a function of $\delta$ in Fig.~\ref{fig:capture_relic_masses} for different dark matter masses, considering $Y=1/2$; the rate is the same for every multiplet with this hypercharge and for both scalars and fermions. The capture rate for $Y=1$ multiplets is simply four times larger. 

Two effects drive the capture rate down with increasing dark matter mass. The
first is the number density, $n_\chi\propto1/m_\chi$. The second is kinematic:
a nucleus of mass $m_A\ll m_\chi$ can absorb at most a fraction $\simeq4m_A/m_\chi$
of the dark matter kinetic energy in a single collision, whereas capture requires
the loss of the full asymptotic kinetic energy $\tfrac12m_\chi u^2$, where $u$ is the dark matter speed at infinity. Only halo
particles with $u\lesssim2v_{\rm esc}\sqrt{m_A/m_\chi}$ can therefore be
captured, and since the flux through this low-velocity window scales as
$u_{\rm max}^2$, the rate acquires a second factor of $1/m_\chi$. For
$m_\chi\gg m_A$ the curves in Fig.~\ref{fig:capture_relic_masses} are simply shifted
copies of one another with $\mathcal{C}\propto m_\chi^{-2}$, and the kinematic
ceiling, set by $\tfrac12\mu_Av^2$ with $\mu_A\to m_A$, is the same for all of
them. The $300$~GeV curve is the exception: its reduced mass with the heaviest
solar nuclei has not yet saturated ($\mu_{\rm Pb}\simeq0.6\,m_{\rm Pb}$), so its
ceiling on each element is lower, and although it starts highest it falls below
the heavier-mass curves for $\delta\gtrsim1.2$~MeV, where capture proceeds only
on the heaviest elements. To reiterate, at the capture step, all the $Y=\frac{1}{2}$ multiplets have identical capture cross sections irrespective of their spin, and the $Y=1$ multiplets just differ by a factor of 4. 

\subsection{Energy loss}
\label{sec:energyloss}
After capture, DM continues to lose orbital energy through the tree-level inelastic transition. As the orbit shrinks, the maximum DM velocity inside the Sun decreases until the inelastic transition becomes forbidden. The deepest accessible orbit $r_{\rm stall}$ is determined by the heaviest solar nucleus, uranium, such that
\begin{equation}
v_{\rm esc}^2(0)
-
v_{\rm esc}^2(r_{\rm stall})
=
\frac{2\delta}{\mu_U}.
\label{eq:rstall}
\end{equation}
Below $r_{\rm stall}$, orbital energy loss must proceed through elastic scattering. The differential
cross section for these collisions can be written as
\begin{equation}
\frac{d\sigma_A}{dq^2}
=
\frac{
A^2 \sigma_{\rm SI} F_A^2(q)
+\sigma_{\rm SD}
}
{4\mu_n^2 v^2},
\label{eq:dsigma_general}
\end{equation}
where $\sigma_{\rm SI}$ and $\sigma_{\rm SD}$ are the per-nucleon cross
sections. The spin-independent term is coherently enhanced and dominated
by the heavy elements; the spin-dependent term carries no $A^2$ enhancement and,
in the Sun, acts essentially only on hydrogen, the one abundant nucleus with
spin. A recoil \(E_R=q^2/(2m_A)\) removes a fraction
\(E_R/E_\chi\) of the DM kinetic energy, with
\(E_\chi=m_\chi v^2/2\). We therefore define the slowing
cross section
\begin{equation}
\sigma_A^{\rm slow}(v)
\equiv
\int_0^{q_{\max}^2}
\frac{E_R}{E_\chi}
\frac{d\sigma_A}{dq^2}\,dq^2,
\label{eq:sigma_slow}
\end{equation}
which gives the orbit-averaged energy-loss rate
\begin{align}
\left\langle\frac{dE}{dt}\right\rangle
=
-\frac{2}{T(r_{\max})}
\int_0^{r_{\max}}dr\,
\sum_A n_A(r)\,
\sigma_A^{\rm slow} \notag\\
\!\left(v(r;r_{\max})\right)v^2(r;r_{\max}).
\label{eq:dEdt_average}
\end{align}
The turning point $r_{\max}$ evolves with $g(r)=GM(r)/r^2$, according to
\begin{equation}
\frac{dr_{\max}}{dt}
=
\frac{1}{g(r_{\max})}
\left\langle\frac{dE}{dt}\right\rangle.
\label{eq:diffeqn}
\end{equation}
For the doublet, we consider two cases. First, the spin-independent loop cross section, consistent with $\sigma_{\rm SI}^{\rm loop}=0$ and second, with the central loop value \(\sigma_{\rm SI}^{\rm loop}\approx 4\times10^{-50}\,\mathrm{cm^2}\). This procedure captures the uncertainty arising from the poorly known $\sigma_{\rm SI}^{\rm loop}$.

In the first case, the thermalization of the fermionic doublet is dominated by SD scattering, while for the scalar doublet we assume that the tree-level inelastic cross section dominates and the orbital evolution stalls at $r_{\rm stall}$. In the second case, thermalization occurs through loop-induced SI scattering for all electroweak multiplets.

\subsection{Evolution of captured dark matter}

We numerically solve Eq.~\eqref{eq:diffeqn} for $r_{\rm max}(t)$ and evaluate it at the present time $t_{\odot}$. Note that this time evolution stalls if $r_{\rm max}$ reaches the thermalization radius
\begin{equation}
R_{\rm therm}
\approx 0.0037\, R_\odot\times \sqrt{\frac{1~\textrm{TeV}}{m_\chi}} .
\label{eq:thermr}
\end{equation}
Following Ref.~\cite{Pospelov:2026ewn}, we estimate the annihilation rate by treating the
captured population as contained within an effective sphere of radius
$R=r_{\rm max}$. The corresponding capture--annihilation
equilibration time is then
\begin{align}
\frac{\tau_{\rm eq}}{t_\odot} &=
\left(\frac{\mathcal{C}}{3\times 10^{19}~\textrm{s}^{-1}}\right)^{-\frac{1}{2}}
\left(\frac{r_{\rm max}
}{0.017\, R_\odot}\right)^{\frac{3}{2}}\nonumber \\
&\quad \times
\left(\frac{1.3 \times 10^{-26}~\textrm{cm}^3\,\textrm{s}^{-1}}
{\langle\sigma v\rangle_{\rm ann}}\right)^{\frac{1}{2}}
\label{eq:teqtsunrmax}
\end{align}
and the present-day annihilation rate corresponds to
\begin{equation}
\Gamma_{\rm ann}(t_\odot)
=
\frac{\mathcal{C}}{2}
\tanh^2\!\left(\frac{t_\odot}{\tau_{\rm eq}}\right).
\label{eq:tanh}
\end{equation}
At equilibrium, i.e., when $\tau_{\rm eq} \ll t_\odot$, $\Gamma_{\rm ann}=\mathcal{C}/2$. 
In practice, for most of the multiplets, comparing Eq.~\eqref{eq:thermr} and Eq.~\eqref{eq:teqtsunrmax} with the appropriate annihilation cross section, we find that thermalization is a sufficient condition for equilibrium for the mass range we consider. It is important to emphasize that complete thermalization is not necessary for achieving the maximum strength of annihilation, as equilibrium typically sets in
earlier. 

\section{Results}
\label{sec:results}
\begin{figure}[t]
    \centering

    \includegraphics[width=1\columnwidth]{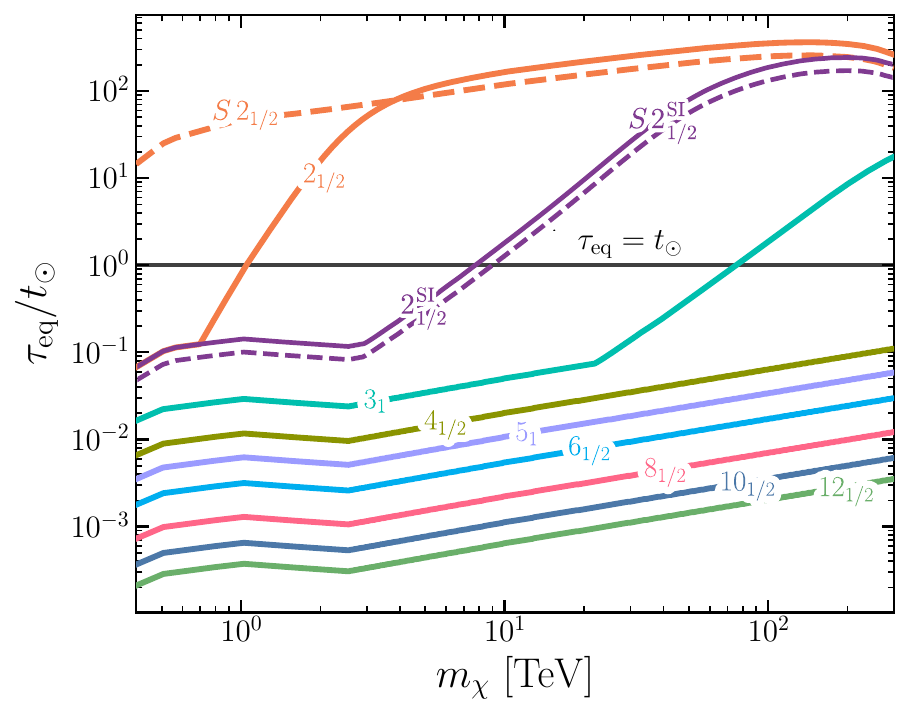}

    \caption{Equilibrium time-scale $\tau_{\rm eq}/t_\odot$ as a function of $m_\chi$. Solid curves show all the fermionic electroweak multiplets, while the dashed orange and purple curves correspond to the scalar doublet under \(\sigma_{\rm SI}^{\rm loop}=0\) and  \(\sigma_{\rm SI}^{\rm loop}=4\times10^{-50}\,\mathrm{cm^2}\), respectively.}

    \label{fig:tau_eq}
\end{figure}
The capture--annihilation equilibration time $\tau_{\rm eq}$ of
Eq.~\eqref{eq:teqtsunrmax} is shown in units of $t_\odot$ as a function of
$m_\chi$ in Fig.~\ref{fig:tau_eq}. It is evaluated at the splitting for which the capture rate saturates
the IceCube limit \(C/2=\Gamma_{\rm IC}^{\rm lim}\). We use $r_{\max}$ from the orbital evolution of
Sec.~\ref{sec:capture}. For $n>2$ only the fermions are shown; the scalars
differ by a factor $\sqrt2$ through the annihilation rate and are otherwise
identical.

\subsection{Higher multiplets}
For $n\ge4$, $\tau_{\rm eq}\ll t_\odot$ over the entire mass range considered.
Both ingredients work in the same direction: the loop-induced elastic cross
section of Eq.~\eqref{eq:sigmaSI_loop} grows as $n^4$ and sinks the population
to the thermal radius well within the solar age, and the annihilation rate of
Table~\ref{tab:nplets} grows with $n$ as well. Capture and annihilation are
therefore in equilibrium, $\Gamma_{\rm ann}=\mathcal{C}/2$, and since the
capture rate depends on the multiplet only through its hypercharge, the solar
constraint in the $(m_\chi,\delta)$ plane is exactly the same curve for every
$n_{1/2}$ with $n\ge4$, for fermions and scalars alike. This is the red curve in
the right panel of Fig.~\ref{fig:npletsmap}.

The triplet is the intermediate case. Its loop cross section is the smallest of
the higher multiplets, $2.9\times10^{-49}~{\rm cm}^2$, and its annihilation rate
the lowest, so that $\tau_{\rm eq}$ rises with mass and crosses $t_\odot$ near
$100$~TeV. Below that mass the $3_1$ shares the equilibrium curve of the $5_1$;
above it, the annihilation rate falls and the constraint weakens. The two are shown as separate
red curves in the left panel of Fig.~\ref{fig:npletsmap}: solid for the $5_1$,
which remains in equilibrium throughout, and dashed for the $3_1$, where it
departs from equilibrium. The departure occurs an order of magnitude above the triplet's
thermal relic mass and does not affect the thermal candidate.

Fig.~\ref{fig:npletsmap} shows the resulting solar constraints together with
the LZ-preferred regions of Sec.~\ref{sec:lz}, for $Y=1$ (left) and
$Y=\tfrac12$ (right), with the thermal relic masses of Table~\ref{tab:nplets}
marked as vertical lines (dashed for fermions, dash-dotted for scalars).
The solar curve excludes everything below it. Its shape follows the capture
rate of Fig.~\ref{fig:capture_relic_masses}: the excluded splitting peaks near
$600$~keV for the $Y=1$ multiplets and $570$~keV for $Y=\tfrac12$ at TeV masses,
and falls with mass as the capture rate drops as $m_\chi^{-2}$, reaching
$\sim350$--$400$~keV at $100$~TeV. The curve is nearly the same for every halo model,
because capture is dominated by the slow part of the velocity distribution to
which the LMC tail contributes nothing; it is the LZ regions that move between
halos.

For the standard halo the LZ-preferred region lies entirely below the solar
curve for every multiplet with a thermal mass below $\sim100$~TeV: the $3_1$
and $5_1$ at $Y=1$, and the $4_{1/2}$ and $6_{1/2}$ at
$Y=\tfrac12$, are excluded as explanations of the event. The $8_{1/2}$ is marginal and only the $10_{1/2}$ and $12_{1/2}$, whose thermal masses lie in the
heavy LZ island where the solar bound has fallen below the preferred
splitting, survive. With the LMC halo, the
preferred splittings move above $500$~keV. For $Y=\frac{1}{2}$, the $4_{1/2}$ becomes
marginal and the $6_{1/2}$ and above become allowed by IceCube constraints. For $Y=1$ the $3_1$ and $5_1$ are no longer excluded by IceCube, but their thermal masses are compatible with the LZ event only at about $3\sigma$. The conclusion is the same for scalars and fermions, whose thermal masses
nearly coincide for $n\ge3$; the two vertical lines for each multiplet in
Fig.~\ref{fig:npletsmap} are barely distinguishable.   

\subsection{Doublets}

The doublets require separate treatment because their loop-induced
spin-independent (SI) cross section is sufficiently small that
uncertainties in its value significantly affect the evolution of
captured dark matter. As discussed in Sec.~\ref{sec:energyloss}, we consider two
benchmarks: $\sigma_{\rm SI}^{\rm loop}=0$, consistent with
higher-order calculations, and
$\sigma_{\rm SI}^{\rm loop}=4\times10^{-50}\,{\rm cm}^2$,
corresponding to its central value from Ref.~\cite{Hill_2014}.

We first consider the limit of vanishing SI scattering. The
fermionic doublet can continue to lose orbital energy through
spin-dependent (SD) scattering on solar hydrogen, as in
Ref.~\cite{Pospelov:2026ewn}. This process is much slower than
the SI scattering of higher multiplets, and the equilibration
timescale is comparable to the solar age at the thermal relic
mass. Since the SD cross section falls as $m_\chi^{-2}$,
further energy loss becomes increasingly inefficient at larger
masses, causing $\tau_{\rm eq}$ to rise steeply, as shown by the solid orange curve in
Fig.~\ref{fig:tau_eq}. The scalar doublet, in contrast, has no
SD interaction. Once inelastic scattering becomes kinematically
forbidden, its orbital evolution stalls at
$r_{\rm stall}\simeq0.1$--$0.2\,R_\odot$, rather than
continuing toward $R_{\rm therm}$. The resulting extended
distribution leads to a much longer equilibration timescale.
At sufficiently large masses, SD scattering becomes ineffective
over the solar lifetime even for the fermionic doublet, so that
both populations remain at approximately $r_{\rm stall}$.
Their equilibration timescales then differ by a factor of
$\sqrt{2}$, reflecting the factor-of-two difference in their
annihilation cross sections.

Including the loop-induced SI interaction changes this
picture. Both doublets can continue to lose energy through
elastic scattering on solar nuclei after the inelastic
channel closes, allowing their orbits to shrink further.
The smaller effective annihilation volume reduces
$\tau_{\rm eq}\propto r_{\max}^{3/2}$, as illustrated by
the purple curves in Fig.~\ref{fig:tau_eq}. The effect is
particularly pronounced for the scalar doublet, whose orbital
evolution would otherwise stall. However, as the dark matter
mass increases, the fractional energy transferred per
collision decreases, making orbital energy loss progressively
less efficient, resulting in $\tau_{\rm eq}$ rising above the
solar age for sufficiently heavy masses both doublets.

The resulting solar constraints are shown in Fig.~\ref{fig:dirac_nplets_shm_lmc}.
For the fermionic doublet, neglecting SI scattering 
excludes splittings $\delta\lesssim566$ keV at the
thermal relic mass of 1.08 TeV. This excludes the
SHM-preferred interpretation of the LZ event, while
leaving only the upper edge of the LMC-preferred region.
Including SI scattering changes the constraint little
at the thermal mass, but substantially strengthens it
at larger masses, excluding splittings below approximately
500 keV at 10 TeV and maintaining sensitivity above
300 keV out to roughly 35 TeV.

For the scalar doublet, the absence of SD scattering
weakens the solar constraint when
$\sigma_{\rm SI}^{\rm loop}=0$. At its thermal relic
mass of 0.58 TeV, splittings below approximately
510 keV are excluded, disfavoring the SHM interpretation
while leaving the LMC interpretation marginally viable.
Including SI scattering raises this bound to approximately
566 keV, excluding the LMC-preferred region as well.
The exclusion also extends to higher masses, remaining
above 300 keV out to roughly 40 TeV, compared with
approximately 20 TeV in the absence of SI scattering.

Thus, unlike the higher multiplets, the doublet limits
depend appreciably on the elastic scattering channel,
particularly for the scalar. At masses above a few tens
of TeV, the solar bounds fall below the splitting range
shown in Fig.~\ref{fig:dirac_nplets_shm_lmc}. Explaining the high-mass LZ-preferred
region with either doublet would in any case require a
non-thermal production history. It is important to point out that a tree-level Higgs
coupling for the scalar, if present, could further
enhance orbital energy loss and strengthen the solar
constraint.

\section{Discussion and conclusions}
\label{sec:conclusions}

Extensive experimental, theoretical and numerical work has made it possible to
probe a large number of well-motivated dark matter models. Electroweak dark
matter is economical in its assumptions, requiring no additional mediator
particles, and definite in its predictions: specifying the representation fixes
the thermal mass, provided the WIMP saturates the dark matter abundance. Its
low-energy phenomenology, and in particular the rate of scattering of Galactic
dark matter on nuclei, depends very sensitively on the mass splitting $\delta$,
which controls whether the tree-level $Z$-mediated cross section is kinematically
accessible. A relatively small change in $\delta$ can change the scattering rate
by many orders of magnitude, with larger $\delta$ favoring smaller rates and
larger recoil energies. This is the pattern displayed by the single event recently reported by the LZ collaboration~\cite{LZ:2026axp}: a very rare event at a recoil energy far above the expectation for elastic scattering. The event can be fit
by electroweak multiplet dark matter with splittings in the range
$\delta\sim300$--$600$~keV. In this paper we have determined
the preferred region of the $\{m_\chi,\delta\}$ parameter space for complex WIMPs. While the
electroweak doublet is somewhat special, all higher representations share the
same preferred mass and splitting once the hypercharge $Y$ is fixed.

The bulk of the paper is an analysis of solar capture and of the resulting
neutrino signal from electroweak dark matter, extending the Higgsino results of
Ref.~\cite{Pospelov:2026ewn} to higher multiplets. We find that the
non-observation of high-energy neutrinos from the Sun by the IceCube
collaboration~\cite{IceCube:2025fcu} places significant constraints on
electroweak dark matter in general. Once captured, dark matter in a higher
representation loses energy and reaches capture--annihilation equilibrium more
quickly than the Higgsino. At the same time, higher representations have higher
thermal masses, which lowers the capture efficiency and hence the neutrino flux.
With the Standard Halo Model as input for the velocity distribution, we find
that all thermal electroweak WIMPs with $Y=1$ are disfavored as explanations of
the LZ event. For $Y=1/2$ one has to go to very high representations, the
$10_{1/2}$ and $12_{1/2}$, to reconcile thermal production, the LZ signal and the
IceCube constraint. Not surprisingly, including the LMC-induced high-velocity
tail in the velocity distribution opens up the parameter space: it raises
the splitting preferred by the LZ event while leaving the solar constraint
essentially unchanged, and many more models become allowed. A detailed understanding of the
high-velocity tail of the WIMP velocity distribution~\cite{OHare:2026nqi} is therefore essential for
the electroweak interpretation of the LZ event. We emphasize the high degree of
universality of the solar constraints in the $\{m_\chi,\delta\}$ plane: apart
from the dependence on $Y$, all higher-representation multiplets predict almost
the same solar signal once the mass and the splitting are fixed.

The IceCube collaboration, and in the future other high-energy neutrino
observatories~\cite{KM3Net:2016zxf,TRIDENT:2022hql,P-ONE:2020ljt}, could improve on these
constraints. Electroweak dark matter is predictive enough that the collaborations
themselves can analyze the inelastic capture and the energy loss of the WIMPs in
the solar interior, inject the appropriate mixture of $W^+W^-$ and $ZZ$ final
states, and account for ``peripheral'' annihilation, outside the solar core,
which suffers less from neutrino absorption. The case of inelastic electroweak
dark matter is prominent enough to be treated on par with the generic
spin-dependent and spin-independent scattering already addressed by IceCube. As
this work shows, inelastic dark matter is a case in which neutrino telescopes
can outperform direct detection, and such analyses would be welcome regardless
of the future status of the LZ event.

Interestingly, the higher electroweak multiplets also present an opportunity for future searches at LZ itself. Unlike the doublet, whose loop-induced spin-independent cross section is close to the neutrino floor and can be consistent with zero, the higher representations generally have appreciable elastic scattering loop cross sections, as seen in Table~\ref{tab:nplets}, with many lying well above the neutrino floor.  Substantially increased exposures could allow LZ or future xenon experiments to probe these candidates through elastic nuclear recoils.

\textit{\textbf{Acknowledgments.}}
H.R. is supported in part by NSF
Grant No.~PHY-2515007, the University
of Delaware Research Foundation and the John Templeton Foundation Award No.~63595. M.P. is supported by the Department of Energy under Grant No.~DE-SC0011842 at the University of Minnesota.

\bibliographystyle{apsrev4-2}
\bibliography{reference.bib}

\end{document}